\documentclass[english,twocolumn,prl]{revtex4}
\usepackage[utf8]{inputenc}
\usepackage{amsmath}
\usepackage{graphicx}
\usepackage[english]{babel}
\usepackage{color}
\usepackage{ulem}
\usepackage[english]{babel}
\usepackage[T1]{fontenc}
\usepackage{lmodern}
\usepackage{amsmath}
\usepackage{amssymb}
\usepackage[amssymb,pstricks]{SIunits}
\usepackage{xspace}
\usepackage{color}
\usepackage{graphics}
\usepackage{epsfig}

\newcommand{\beq}{\begin{equation}}
\newcommand{\eeq}{\end{equation}}

\newcommand{\ket}[1]{\ensuremath{|#1\rangle}\xspace}
\newcommand{\bra}[1]{\ensuremath{\langle #1|}\xspace}
\newcommand{\braket}[1]{\ensuremath{\langle #1 \rangle}\xspace}
\newcommand{\dd}{\mathrm{d}}

\newcommand{\tg}[1]{\textcolor{blue}{#1}}

\usepackage{babel}

\begin{document}

\title{Phonon-tuned bright single-photon source}

\author{Simone Luca Portalupi$^{1}$}
\author{Gaston Hornecker$^{2}$}
\author{Val\'erian Giesz$^{1}$}
\author{Thomas Grange$^{2}$}
\author{Aristide Lema\^itre$^{1}$}
\author{Justin Demory$^{1}$}
\author{Isabelle Sagnes $^{1}$}
\author{Norberto D. Lanzillotti-Kimura$^{1}$}
\author{Lo\"ic Lanco$^{1,3}$}
\author{Alexia Auff\`eves$^{2}$}\email{alexia.auffeves@grenoble.cnrs.fr}
\author{Pascale Senellart$^{1,4}$}\email{pascale.senellart@lpn.cnrs.fr}


\affiliation{$^{1}$ CNRS-LPN Laboratoire de Photonique et de Nanostructures, Route de Nozay,  91460 Marcoussis, France}

\affiliation{$^{2}$ CEA/CNRS/UJF joint team ``Nanophysics and Semiconductors'', Institut N\'eel-CNRS, BP 166, 25 rue des Martyrs, 38042 Grenoble Cedex 9, France}

\affiliation{$^{3}$ Département de Physique, Université Paris Diderot, 4 rue Elsa Morante, 75013 Paris, France}

\affiliation{$^{4}$ Département de Physique, Ecole Polytechnique, F-91128 Palaiseau, France}

\begin{abstract}
{Pure and bright single photon sources have recently been obtained by inserting  solid-state emitters  in photonic nanowires  or microcavities.
The cavity approach presents the attractive possibility to greatly increase the source operation frequency. However, it is perceived as technologically demanding because the emitter resonance must match the cavity resonance.
Here we show that the spectral matching requirement is actually strongly lifted by the intrinsic coupling of the emitter to its environment. A single photon source consisting of a single InGaAs quantum dot inserted in a micropillar cavity is studied. Phonon coupling  results in a large Purcell effect even when the quantum dot is detuned from the cavity resonance. The phonon-assisted cavity enhanced  emission is shown to be a good single-photon source, with a brightness exceeding $40$ \% for a detuning range covering $15$ cavity linewidths. }
\end{abstract}

\maketitle

Atomic like emitters such as semiconductor quantum dots (QDs), defects in diamond  or colloidal nanocrystals  are solid-state single\tg{-}photon sources \cite{ michler2000,beveratos2002,michler2000NC} which have potential applications in quantum cryptography and metrology. While the quantum nature of the emitted light has long been established in each system, the main challenge for applications is to efficiently collect the emitted photons. This emission is almost isotropic if no engineering of the electromagnetic field surrounding the emitter is used. Several technological approaches have allowed very high extraction efficiencies recently: inserting the emitter in a dielectric antenna like a photonic nanowire \cite{claudon2010,lukin2010} or  in an optical microcavity \cite{gazzanonatcom2013,dousse2010}. The latter approach, based on cavity quantum electrodynamics, has been shown to be a powerful tool to control both the emission pattern and emission rate of a single emitter \cite{gayralgerard}. By coupling the emitter to the confined optical mode of a microcavity, the emission rate into this mode is enhanced by the Purcell effect \cite{Purcell,gerard1998}. This property has been successfully used to fabricate ultrabright single photon sources \cite{gazzanonatcom2013,dousse2010,gazzanoPRL2013}. Indeed, by increasing the emission rate into a chosen mode of the electromagnetic field by the Purcell factor $F_P$, a fraction $\beta=F_P/(F_P+1)$ of the photons are emitted into the cavity mode, provided that the emission rate in any other mode is not significantly affected. High Purcell factors not only allow efficient collection of the photons but also increase the operation rate of the source, by reducing the radiative decay process hence the time interval between consecutive excitation pulses \cite{highfrequency}. The Purcell effect, however, requires that the optical transition of the emitter  is spectrally matched to the resonance of the cavity mode . Such condition is difficult to obtain with solid-state emitters due to the large dispersions in their emission frequencies.

 Solid-state emitters also strongly interact with their solid state environment: phonon-assisted emission is observed in every system, representing few percents of the overall emission for QDs  \cite{favero2003,besombes2002,peter2004} but around $90$ \% for defects in diamond \cite{phononNV}. Such interaction  is usually seen as a drawback for quantum applications. Here we show that this strong interaction  with the environment is actually an efficient tool to obtain cavity-based bright single-photon sources: it provides  a built-in spectral tuning to the cavity resonance. We demonstrate a phonon-tuned bright single-photon source for a single QD  coupled to a pillar cavity.

\begin{figure}[h!]
	\begin{center}
		\includegraphics[width=1.00\linewidth]{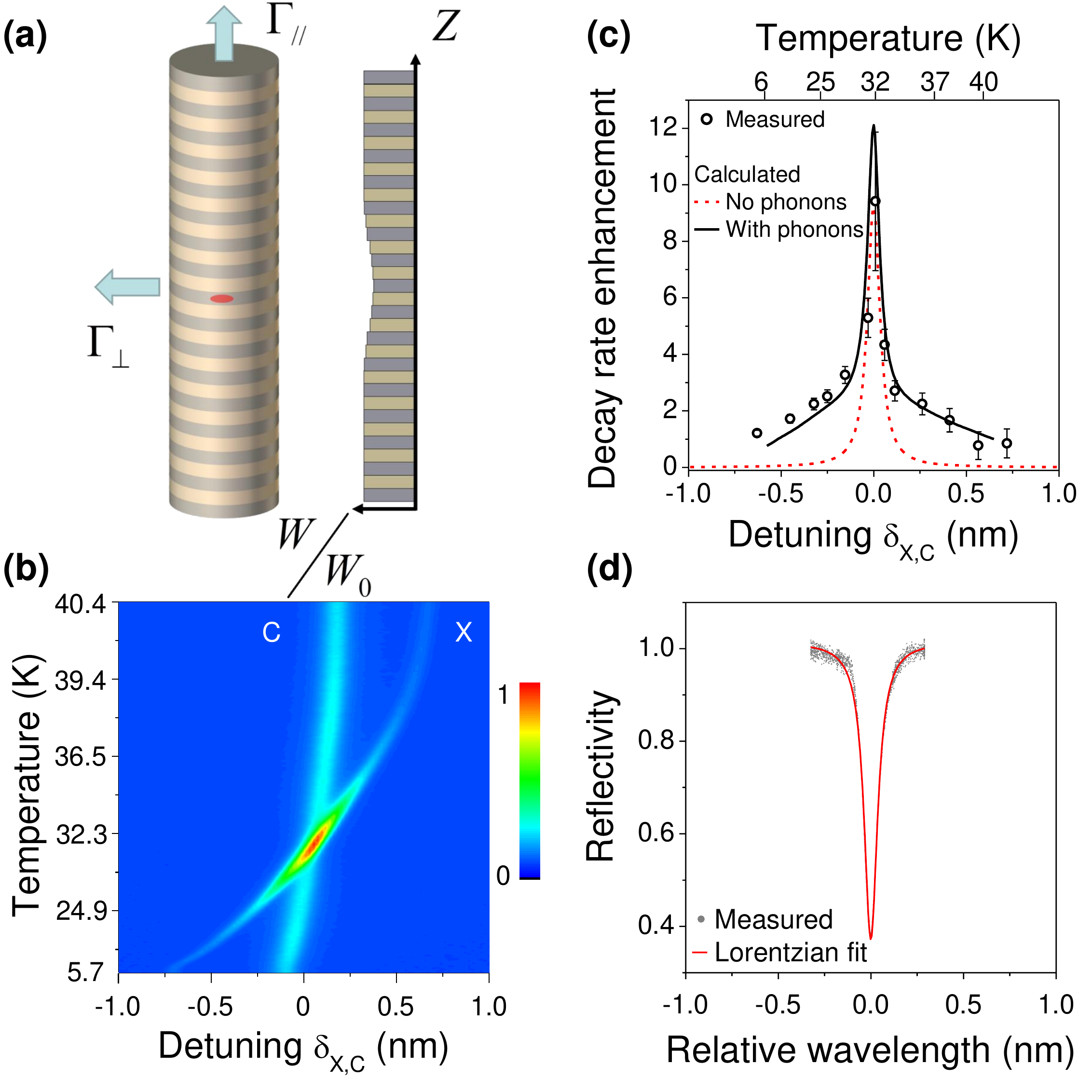}
	\caption{(a) schematic of the system under study: a single quantum dot deterministically positioned in a micropillar cavity. The right plot presents the layer thickness $W$ normalized to the nominal $W_0=\lambda/4n$ thickness (supplementary). The arrows indicate the radiative decay rate into the cavity mode $\Gamma_{\parallel}$ and into the other modes $\Gamma_{\perp} $  (b) Emission intensity of the device as a function of temperature and energy. The QD exciton line is resonant to the cavity line at $32 \ K$. (c) Measured radiative decay time normalized to the decay time in the planar cavity $\tau_X^{0}/\tau_X - 1$ as a function of temperature (top) and detuning $\delta_{X,C}$ (bottom). (d) Reflectivity spectrum of the pillar under study, showing the cavity mode with a FWHM $\gamma_c=80 \ pm$.}
	\label{fig:1}
\end{center}
\end{figure}

We study a single InGaAs quantum dot positioned at the maximum of a pillar cavity fundamental mode with $50 \ nm$ accuracy by means of the in-situ lithography technique \cite{dousse2008}. The vertical structure for the cavity presents an adiabatic design of mirror layers that allows minimizing sidewall losses \cite{adia}(figure 1a and supplemental). As shown below, this allows reaching high extraction efficiencies  with high Purcell factors. 
Electron hole pairs in QDs couple to acoustic phonons: phonon sidebands are observed in emission spectroscopy \cite{favero2003,besombes2002,peter2004}, resulting from the emission of a photon together with the emission or absorption of one or few phonons. In InGaAs QDs, phonon sidebands represent typically few percents of the total emission at 10 K, yet they lead to new effects for  QDs  inserted in cavities \cite{finley2007,valente2014}.
The first effect of  phonon coupling is to lead to acceleration of spontaneous emission when the QD is spectrally detuned from the cavity mode. Such effect, first evidenced for QDs in photonic crystal cavities \cite{finley2007,lodahl2012}, is also observed here in micropillar cavities.  In figure 1b, the emission intensity of the QD-pillar under study is presented as a function of temperature and wavelength. For each temperature, two emissions lines are observed, corresponding to the emission of the QD exciton (X) resonance and the emission within the cavity mode resonance (C). Adjusting the temperature allows changing the X-cavity detuning  $\delta_{X,C}=\lambda_X-\lambda_{C}$ between the cavity and the QD X resonances.  Figure 1c presents the measured enhancement of the emission decay rate into the cavity mode, namely $F_P^{eff}(\delta_{X,C})=\frac{\tau_X^{0}}{\tau_X(\delta_{X,C})}-1$ with $\tau_X$ the measured decay rate of the X line and $\tau_X^{0}=1.3\ ns$ the X decay rate in the planar cavity. The top axis presents the sample temperature and the bottom axis $\delta_{X,C}$. A strong enhancement of decay rate is observed, reaching $10\pm2$ at resonance. For an ideal two-level system coupled to a cavity mode, the acceleration of spontaneous emission versus $\delta$ follows a Lorentzian dependence $F_P(\delta_{X,C})=F_P \frac{\gamma_c^2}{\gamma_c^2+4\delta_{X,C}^2}$ (dashed curve in fig 1c) where  $\gamma_c =80 \ pm$ is the cavity linewidth (corresponding to a Q factor of Q=12000), as measured in reflectivity (figure 1d). The experimentally measured acceleration of spontaneous emission (symbols in fig 1c) does not follow this simple lorentzian dependence, but evidences the imprint of phonon-assisted mechanisms, with enhanced acceleration of spontaneous emission outside the cavity resonance.

\begin{figure}[h!]
	\begin{center}
		\includegraphics[width=1.00\linewidth]{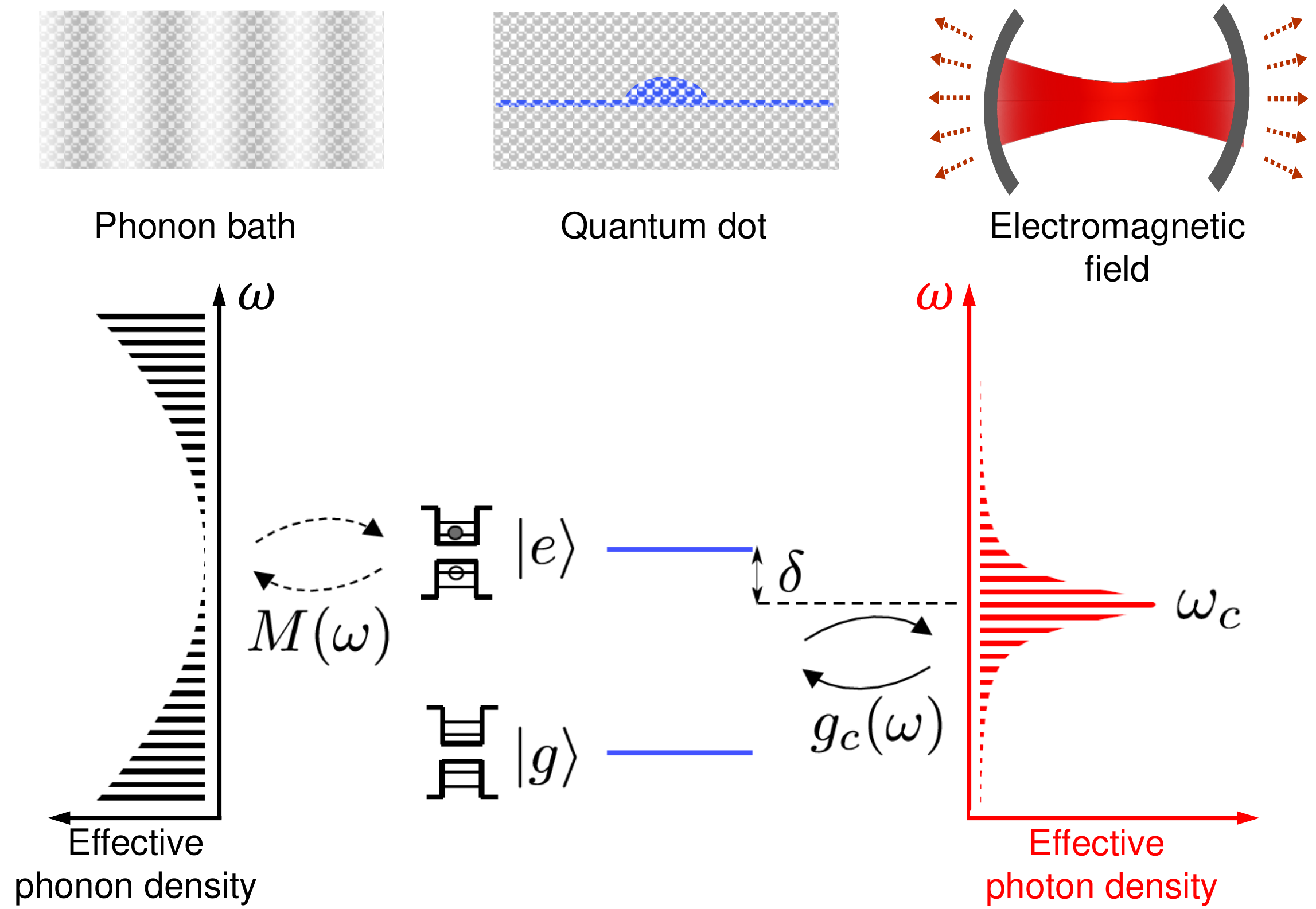}
	\caption{The quantum dot is considered as a two-level system and the phonon bath and electromagnetic field are described as continua. The density of states  of the electromagnetic field is modified by the cavity and is peaked about its resonance frequency $\omega_c$. The coupling to the phonons is treated non-perturbatively using the independent bosons model while the coupling to the electromagnetic field is treated in first order perturbation theory (see text and supplementary).}
	\label{fig:2}
\end{center}
\end{figure}

The influence of phonon-assisted mechanisms on the Purcell effect is modeled by considering the system sketched in figure 2.  A two-level system is coupled to a longitudinal acoustic phonon bath and to the electromagnetic field. The density of states of the latter is modified by the cavity and peaked around its resonance frequency $\omega_c$, detuned from the QD resonance by $\delta_{X,C}$. The coupling to the phonons is described using the independent bosons model \cite{mahan} while the photon emission is treated in first--order perturbation theory, in the weak coupling regime (see supplementary). The QD exciton state can radiatively recombine at its intrinsic frequency $\omega_{X}$ with no associated phonon  process (zero phonon line, ZPL) or at a different energy by emitting or absorbing one or several phonons (phonon sideband, PSB). Eventually, both the zero phonon line and the phonon assisted emission channels experience Purcell effect leading to enhanced decay rate into the cavity mode $\Gamma_{\parallel}^{ZPL} $ and $\Gamma_{\parallel}^{PSB}$.  Phonon coupling thus leads to an enhanced total exciton decay rate $\Gamma_{tot}=\Gamma_{\parallel}^{ZPL}+\Gamma_{\parallel}^{PSB}+\Gamma_{\perp}$  on a large detuning range. We  introduce here $\Gamma_{\perp}$, the emission decay rate for both ZPL and PBS emission mechanisms  into the other optical modes of the electromagnetic field, which  is close to the emission decay rate in bulk material \cite{gerard1998}. The calculated effective Purcell factor, defined as $F_P^{eff}(\delta_{X,C})=\frac{\Gamma_{\parallel}^{ZPL}+\Gamma_{\parallel}^{PSB}}{\Gamma_{\perp}}$ is shown in fig 1c (solid line): it reproduces well the experimental observations, showing the strong influence of phonon mechanisms on the QD coupling to the cavity mode. 

These phonon assisted mechanisms also have a distinctive signature in the emission spectrum of the device. Figure 3a presents the calculated emission spectrum $S(\lambda)$ for a single QD state coupled to the cavity mode . It is the product of the cavity  spectrum $S_{C}(\lambda)$ (red line) and the QD emission spectrum in bulk $S_{QD} (\lambda)= S_{QD}^{ZPL}+ S_{QD}^{PSB}$ (blue line) where the two terms correspond to the zero-phonon line and the  phonon sidebands \cite{valente2014}. The total emission spectrum (black line) thus presents two emission peaks, corresponding to either the ZPL or the PSB coupled to the cavity. Because the PSB spreads over a few nanometers spectral range, thus largely overcoming the cavity width $\gamma_c=80 \ pm$, the second term gives rise to an emission peak roughly centered at the cavity resonance \cite{valente2014}. This emission line is the cavity line (C) and results from the phonon sideband emission enhanced by the cavity.   Figure 3b, presents  the measured emission spectrum at $12 \ K$ for an excitation power close to saturation of the exciton line. Three emission lines are observed: the exciton X, the biexciton XX and an intense cavity line C. To unravel the nature of each emission line, photon correlation measurements are performed ( Figure 3c). The top curve shows the exciton autocorrelation  function $g_{X,X}^{(2)}(\tau)$ and evidences a good single photon purity with $g_{X,X}^{(2)}(0)=0.03\pm0.02$.  Strikingly, cross correlation of the X and C lines also shows a strong  antibunching $g_{C,X}^{(2)}(0)=0.15$: this observation shows  that the emission in the cavity mostly arises from the X phonon sideband. The middle curve finally shows the autocorrelation curve for the cavity line: a strong antibunching with $g_{C,C}^{(2)}(0)=0.25\pm0.05$ is observed showing that the cavity line is  a very good single photon source. 

\begin{figure}[h!]
	\begin{center}
			\includegraphics[width=1.00\linewidth]{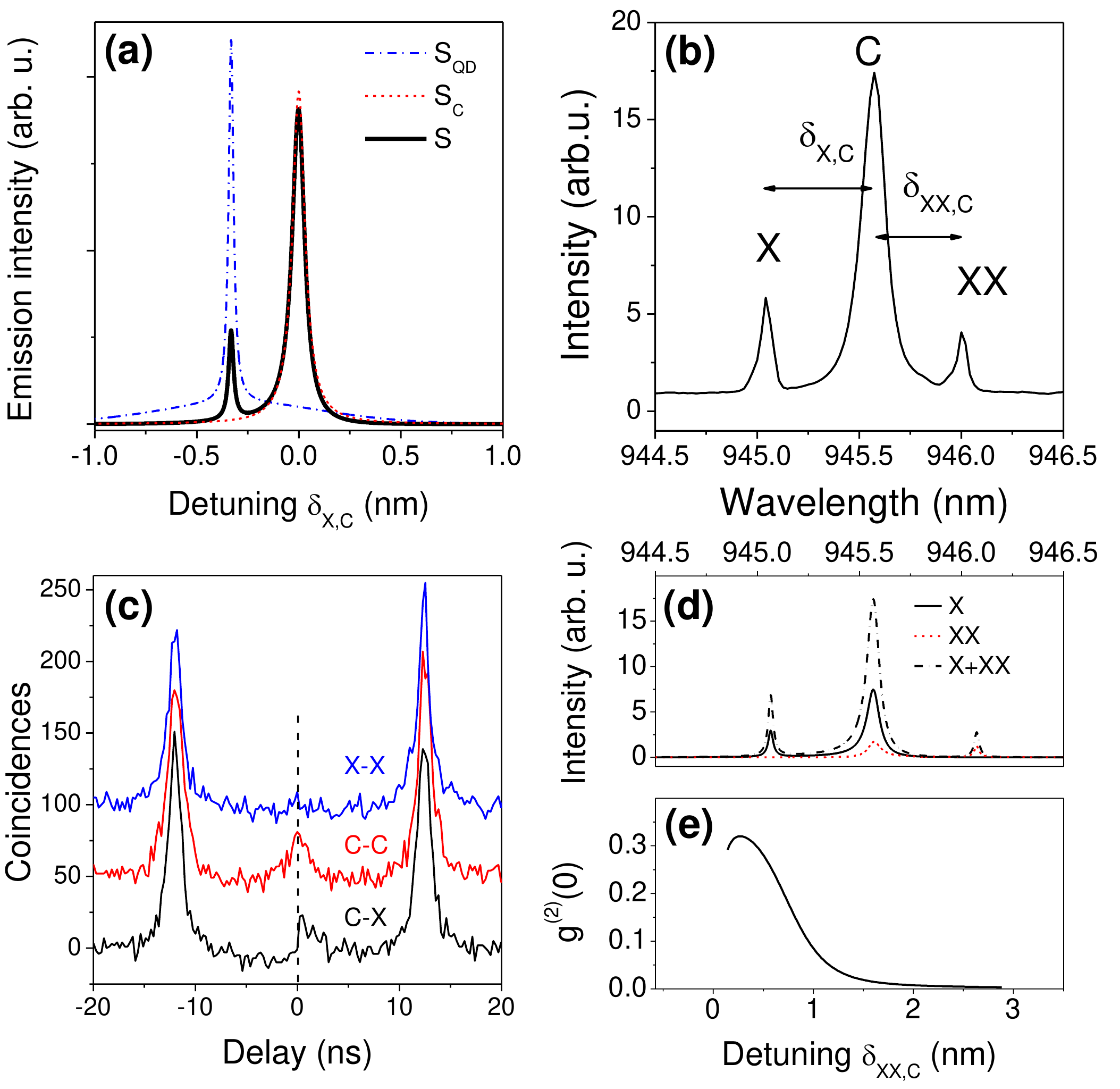}\caption{ (a) Black: calculated emission spectrum $S(\lambda)$ for a QD exciton line coupled to cavity line, with  $ \delta_{X,C}=- 0.5 \ nm$. The blue line is the QD spectrum $S_{QD}(\lambda)$ of the QD coupled to the phonon bath but not coupled to the cavity; the red line is the cavity spectrum $S_{C}(\omega)$. (b)  Emission spectrum measured at $12$ K at an excitation power close to the saturation.  (c) Photon correlation measurements: exciton autocorrelation (blue); cavity line autocorrelation (red); exciton-cavity cross correlation (black). Experimental curves have been vertically shifted for clarity. (d) Calculated total spectrum (dashed line) with the contributions from the X line (solid black) and XX line (dotted red) to the cavity emission.  (e) Calculated values of $g_{C,C}^{(2)}(0)$ for different cavity biexciton detunings. The X-C blue detuning is fixed to $\delta_{X,C}=-0.5 \ nm$ and the temperature is set to $12$ K. The pillar under study shows a $\delta_{XX,C} = 0.5\ nm$.}
		\label{fig:3}
	\end{center}
\end{figure}

The brightness of this phonon-tuned single-photon source is measured by  carefully calibrating the experimental setup under a pulsed excitation at $82 \ MHz$ (see supplementary). The brightness $B$, defined as the number of photons collected per pulse in the collection lens is derived from the number of counts on the detector $I$, using $B= I \ \eta \sqrt{1-g^{(2)}(0)} $ where $\eta=1.44 \pm 0.1 \%$ is the overall setup detection efficiency  and the last term is the correction from multiphoton events. The top panel of figure 4a presents the brightness  of various lines as a function of power (top panel), with the corresponding  $g^{(2)}(0)$ (bottom panel) for two detunings. 
 The black symbols present the case where the X is resonant to the cavity mode ($\delta_{X,C} = 0$). At saturation, the corrected brightness amounts to $B_X(\delta_{X,C}=0)=74\pm7\%$ collected photon per pulse, very close to record brightnesses of $78 \pm8 \%$ recently reported \cite{gazzanonatcom2013}. The red squares present the brightness and $g_{C,C}^{(2)}(0)$ for the cavity line at $12 \ K$, when the X line is blue detuned from the cavity line by $\delta_{X,C}=-0.5\ nm$. At saturation, a brightness  of $B_C(\delta_{X,C}=-0.5\ nm)=35\pm 3\%$ is obtained. The brightness of the detuned X line   for the same detuning reaches $B_X(\delta_{X,C}=-0.5\ nm)=5$ \% at saturation (red triangles). As a result, a total brightness  $B_X(\delta_{X,C}=-0.5\ nm)+B_C(\delta_{X,C}=-0.5\ nm)$   as large as $40\pm4\%$ (red crosses) is obtained for a detuning representing 6 cavity linewidths.

\begin{figure}[h!]
	\begin{center}
		\includegraphics[width=1.00\linewidth]{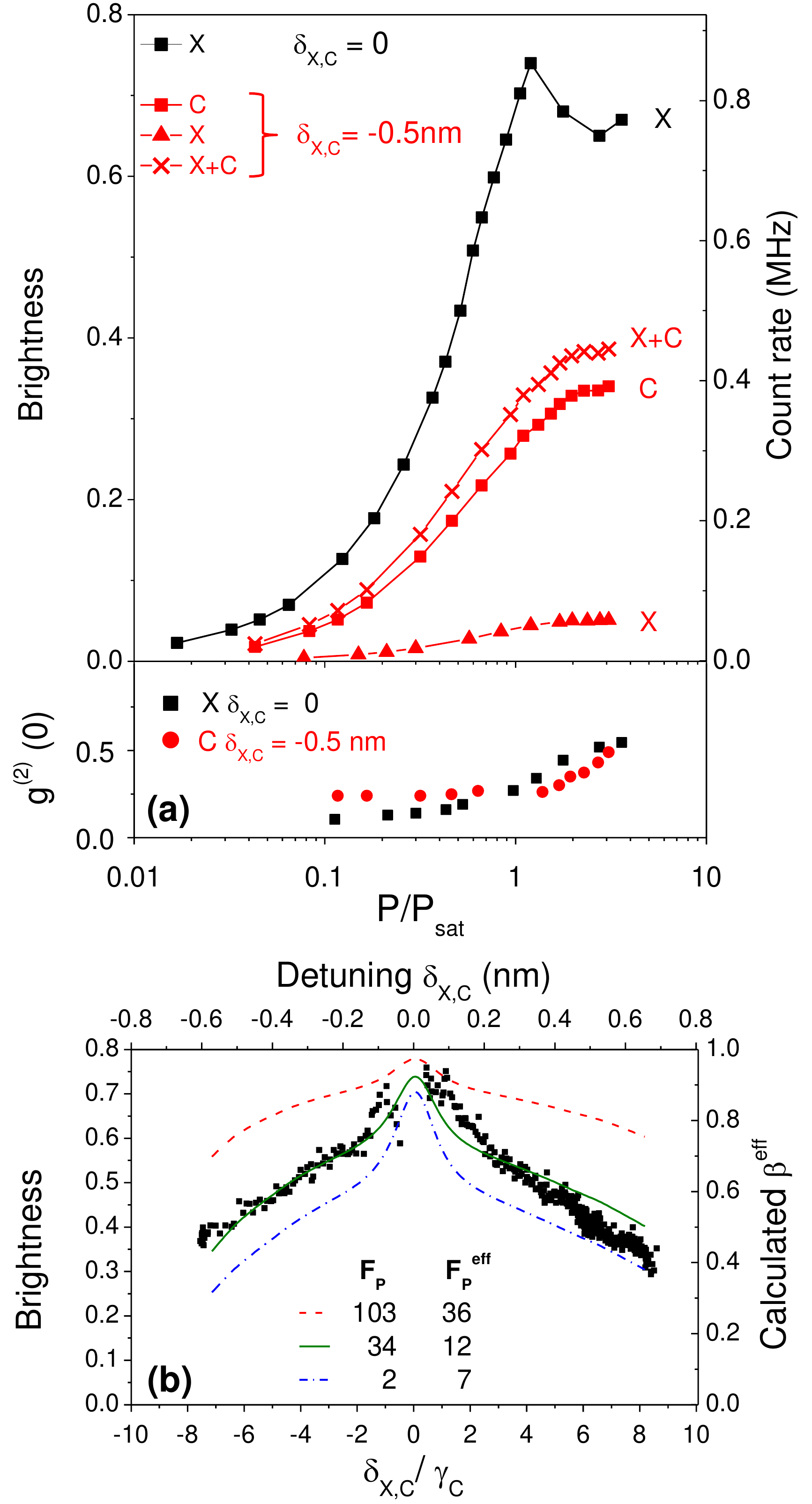}
		\caption{(a) Top: collected photons per pulse for different lines. Black squares: X line for $\delta_{X,C}=0$. Red symbols are measured for non-zero detuning ($\delta_{X,C}= - 0.5 \ nm$). Red squares:  cavity line. Red triangles: X line. Red crosses: total brightness, sum of C and X lines. Bottom: corresponding measured  $g^{(2)}(0)$ for X at $\delta_{X,C}=0$ (black squares) and C at $\delta_{X,C}= - 0.5 \ nm$ (red circles).
		(b) Total brightness $B =B_X+ B_{C}$ (black squares, left scale) compared with calculated mode coupling $\beta$ (right scale) as a function of the detuning normalized to the cavity linewidth ($\delta_{X,C}/\gamma_C$) for different values of $F_P$.}
		\label{fig:4}
	\end{center}
\end{figure}

The brightness of the source is now studied as a function of the detuning. The total  measured brightness $B (\delta_{X,C})=B_X(\delta_{X,C})+ B_{C}(\delta_{X,C})$ is plotted on figure 4b: brightness exceeding $40 \pm 3\%$ are demonstrated over a $[-0.65\ nm, 0.65 \ nm]$ detuning range.  As explained in \cite{gazzanonatcom2013}, the source brightness is given by $B=\beta^{eff} \frac{Q}{Q_0}q_X$ where $\beta^{eff}$ is the fraction of emission into the mode, the $Q/Q_0$ term accounts for the sidewall losses ($Q_0$ being the planar cavity quality factor) and $q_X$ is the QD X state occupation factor at saturation.
In the presence of phonon coupling $\beta^{eff}=\frac{\Gamma_{\parallel}^{ZPL}+\Gamma_{\parallel}^{PBS}}{\Gamma_{tot}}$ results from emission into the ZPL and PBS emission into the cavity mode. $\beta^{eff}$ is plotted on the right scale of figure 4b for various nominal Purcell factors $F_P$. The corresponding effective Purcell factors at resonance (32 K) are also indicated. To compare the measured brightness  to the model, we deduce $\frac{Q}{Q_0}q_X$ close to zero detuning. The  experimental observations are well reproduced considering an effective Purcell factor around $12$, a figure consistent with the effective  Purcell factor measured at resonance, within error bars. This calculation further shows that the larger the Purcell factor, the broader the wavelength range where the device acts as a very bright single photon source. For a  nominal (effective) Purcell factor of 103 ($36$), a brightness exceeding $60$ \% can be reached in a detuning range covering $15$ times the cavity linewidth.

 The purity of the resulting single-photon source presented here is good with $g_{C,C}^{(2)}(\tau)\approx 0.2$ at 12 K, but it could be further improved as discussed below. Careful analysis of  the zero delay peak in $g_{C,X}^{(2)}(\tau)$ in figure 3c shows a strong temporal asymmetry, with the main signal for positive delays. Moreover, the area of the central peak is  roughly half that of the cavity autocorrelation curve $g_{C,C}^{(2)}(\tau)$. This observation is the signature of the radiative cascade between the XX and X lines. When several optical transitions of the QD are coupled to the cavity mode, the phonon sidebands of the various transitions feed the cavity line.  However, because the XX line at $12 \ K$ is on the low energy side of the cavity mode, the phonon processes resonant with the cavity arise from the absorption of a phonon, a process less efficient than phonon emission. Figure 3d shows the calculated contributions of various phonon sidebands to the total spectrum. From these spectra, the calculated $g_{C,C}^{(2)}(0)$ is estimated to be $0.27$ (figure 3d), close to the experimental value. Here the X to XX spectral separation for this QD is  small ($<1 \ nm$). Larger X-XX binding energies are commonly observed for InAs QDs  as well as in II-VI QDs, for which pillar cavity technology is well established \cite{tomaz}. The purity of the source can be greatly enhanced for QDs presenting higher XX binding energy, as shown in figure 3d presenting the calculated $g_{C,C}^{(2)}$ of the cavity emission line at $12K$ for increasing biexciton-cavity detuning.

Our results show that the solid-state environment can be a resource for the reproducible fabrication of bright quantum light sources, with a wavelength set by the cavity resonance \cite{auffevesPRA2009}. Such concepts developed here for InGaAs QDs can apply to other solid-state emitters, hence strongly reducing the technological constraints. As an illustration, we use the same theoretical formalism and study the case of a NV center in diamond, for which the PSB spreads over $100 \ nm$ and can  represent more than $90$ \% of the emission \cite{phononNV}. For such a NV center inserted in a cavity with Q=250, we show in the supplemental that the coupling to the mode can reach very high values, above $40 \%$, over a very large spectral range as long as the nominal Purcell factor $F_P$ is above $20$, a value already accessible with current technology. Exploitation of phonon assisted mechanisms in the framework of cavity quantum electrodynamics opens the path to room temperature implementations of quantum information protocols.

\vspace{1cm}

\begin{acknowledgments}

{\bf Acknowledgments}

 This work was partially supported by the French ANR QDOM, ANR P3N CAFE, the Fondation Nanosciences de Grenoble, the ERC starting grant 277885 QD-CQED, the CHISTERA project SSQN, the RENATECH French network, the Laboratoire d'Excellence NanoSaclay, the EU-project WASPS. The authors aknowledge D. Valente for help in the theoretical modeling. 
\end{acknowledgments}

\section{Supplementary}

\section{Sample Fabrication}

The sample was grown by molecular beam epitaxy with an adiabatic design \cite{adia}. In order to form the cavity, the thickness $W$ of the GaAs (AlAs) layers is modified adiabatically with respect to the nominal thickness $W_0=\lambda/4n$, decreasing it progressively by: $0.01$, $0.07$ $(0.04)$, $0.15$ $(0.12)$, $0.19$ $(0.18)$, $0.15$ $(0.12)$, $0.07$ $(0.04)$, $0.01$ percent. The central GaAs layer, with decreased thickness of $0.19$ \%, is the one where the InGaAs quantum dot is inserted. 18 (32) pairs of nominal $\lambda/4n$ GaAs/AlAs Bragg mirrors are placed on top (bottom) of this adiabatic cavity.
 The \textit{in-situ} lithography technique was then used  to fabricate micropillar cavities centered within 50 nm from a selected QD, presenting a cavity mode spectrally matched to the X resonance within 1 nm \cite{dousse2008}.

\vspace{0.5cm}

\section{ Setup description, calibration and brightness}

The setup used for the measurements allows us to perform micro-photo\-luminescence, reflectivity and correlation measurements. Light emission is collected from the top of the micropillar with a 0.54 NA aspheric lens. This signal is sent to a Hanbury-Brown and Twiss setup composed by a non-polarizing $50:50$ beam splitter,  two spectrometers equipped with single photons avalanche detectors.  A polarizer and a halfwave plate are placed in the detection line in order to select the emission from one QD axis and to align the incident linear polarization with the direction of maximum efficiency of the spectrometer. The measure of the maximum brightness is derived by taking the sum of the count rates of the two orthogonal QD axes. In order to quantify the brightness of the source all components of the experimental setup have been carefully calibrated giving the results summarized in table I.

 \begin{table}[h!]
	\label{tab:cal}
	\centering
		\begin{tabular}{lcc}
\hline
Component &	Transmission/Efficiency	& Error bar \\
\hline
Detection &	8.8 \%	& $\pm$ 0.5 \% \\
HBT NPBS	& 51 \%	& $\pm$ 2 \% \\
Cryostat window	& 98 \%	& $\pm$ 1 \% \\
Polarizer	& 90 \%	& $\pm$ 2 \% \\
Long pass filter	& 98 \%	& $\pm$ 1 \% \\
Half wave plate	& 98 \%	& $\pm$ 1 \% \\
Collection NPBS	& 40 \%	& $\pm$ 2 \% \\
Lens	& 95 \%	& $\pm$ 1 \% \\
\hline
\end{tabular}
\caption{Transmission and overall efficiency of all setup components. }
\end{table}

Detection is accounting for the response of the full detection system, i.e. spectrometer, in- and out- coupling lenses and single photons avalanche detector efficiency. It is measured by sending a laser signal of known power and repetition rate of $82$ MHz into the detection line. The ratio between the known photon flux and the detected number of counts gives an efficiency of around $8.8 \pm 0.5$ \%. The overall setup efficiency is $\eta=1.44 \pm 0.1 \%$.

\section{Spontaneous emission of a quantum dot coupled to a continuum of phonons in a cavity}

The quantum dot (QD) is modeled as a two-level system coupled to a continuum of longitudinal acoustic (LA) phonons on one hand and to a continuum of electromagnetic (EM) modes on the other hand. The considered EM modes are whether the cavity mode, that leads to an emission in the direction parallel to the pillar axis, or the plane modes, that leads to a perpendicular emission (see Fig.1a in the text). The coupling to the phonons is treated exactly using the independent bosons model. The interaction with the EM field is treated in the rotating wave approximation (RWA) with the Jaynes-Cummings Hamiltonian. The total Hamiltonian of our system is given by

\begin{equation}
H = \hbar\omega_{0}\sigma_+\sigma_- +  H_{\text{ph}} + H_{\text{EM}}  ,
\end{equation}
where $\omega_{0}$ is the frequency of the bare exciton; $\sigma_+=\sigma_-^\dag=\ket{e}\bra{g}$ with $\ket{g}$ and $\ket{e}$ are respectively the ground and excitonic states of the quantum dot; $H_{\text{ph}}$ and $H_{\text{EM}}$ stands for the parts of the Hamiltonian involving  respectively phononic and EM modes.
The phononic part reads
\begin{equation}
 H_{\text{ph}} =  \int d\omega_\mathbf q \rho_{ph}(\omega_\mathbf q) \hbar\omega_\mathbf q b_\mathbf q ^\dag b_\mathbf q
	+ \int d\omega_\mathbf q \rho(\omega_\mathbf q) \hbar M(\omega_\mathbf q) \sigma_+\sigma_-(b_\mathbf q + b_\mathbf q ^\dag),
\end{equation}
where
$b_\mathbf q$ and $\rho_{ph}(\omega_\mathbf q)$ are the ladder operator and the density of states (DOS) of the LA phonon mode of wavevector $\mathbf q$ respectively; $M(\omega_\mathbf q)$ is the deformation potential coupling between this phonon mode and the exciton. We consider a linear dispersion relation for the LA phonons, and spherical wave functions for the electron (hole) with a gaussian shape: $\psi_{e/h}(\mathbf r) = (4\pi\sigma_{e/h}^2)^{-3/4}e^{-\mathbf r ^2/(8\sigma_{e/h}^2)}$. We denote $c_s$ the speed of sound, $\rho_m$ the mass density, $D_{e/h}$ the deformation potential of the electron/hole, and $V$ the volume of the crystal. The DOS and coupling terms read respectively:

\begin{equation}
 \rho_{ph}(\omega_\mathbf q) = \frac{V\omega_ \mathbf q^2}{2\pi^2 c_s^3},
\end{equation}

\begin{equation}
 M(\omega_\mathbf q) = \sqrt{\frac{\omega_\mathbf q}{2\hbar c_s^2\rho_m V}}\cdot\left| D_e e^{-\sigma_e^2\omega^2/c_s^2} + D_h e^{-\sigma_h^2\omega^2/c_s^2} \right|,
\end{equation}

The parameters used are: $c_s = 5000 \,\meter/\second$, $\rho_m = 5320 \,\kilo\gram/\meter^3$, $D_e=D_h=9\,\electronvolt$, $\sigma_e=\sigma_h=3\,\nano\meter$, $V_c=7.6\,\micro\meter^3$, where $D_e, D_h, \sigma_e, \sigma_h$ and $V_c$ have been adjusted to fit the experimental data.\\

On the other hand, the interaction with the EM modes reads:

\begin{multline}
 H_{\text{EM}}  =  \sum_{\nu=\parallel,\perp}  \int d\omega S_\nu(\omega) \hbar\omega_{\nu}a_{\nu}^\dag(\omega) a(\omega)_{\nu}\\
	+  \sum_{\nu=\parallel,\perp} \int d\omega S_\nu(\omega) \hbar g_{\nu}(\omega)(\sigma_+a_{\nu}(\omega) + \sigma_-a_{\nu}^\dag(\omega)) ,
\end{multline}

in which the index $\nu = \parallel, \perp$ stands for the two possible channels for the QD relaxation: either the cavity mode or the plane modes. $a_{\nu}(\omega)$ is the ladder operator for the photon mode of pulsation $\omega$. The Jaynes-Cummings coupling $g_\nu(\omega)$ is given by

\begin{equation}
 g_\nu(\omega) = d\sqrt{\frac{\omega}{2\hbar\epsilon_0V_\nu}},
\end{equation}

\noindent where $d$ is the electric dipole of the quantum dot and $V_\nu$ is either the effective volume of the cavity mode or the diverging quantization volume of the free field. The DOS for the cavity mode is given by a Lorentzian of width $\kappa =  \omega_c/Q = 2\pi c \gamma_C/\lambda_C^2$, with $Q=12000$: 

\begin{equation}
 S_\parallel(\omega) = \frac{1}{\pi} \frac{\kappa/2}{(\kappa/2)^2 + (\omega-\omega_c)^2}.
\end{equation}

\noindent $\omega_c = 2\pi c/\lambda_C$ is the central frequency of the cavity. The DOS for the plane modes is taken as the DOS in free space:

\begin{equation}
 S_\perp(\omega) = \frac{\omega^2V_f}{\pi^2c^3} .
\end{equation}

Emission in both channels is well described in first order perturbation theory, such that the spontaneous emission rate $w_{\nu}(\omega)$ in each mode of pulsation $\omega$ and the total rate per channel $\Gamma_\nu$ are

\begin{equation}
 w_\nu(\omega) = g^2_\nu(\omega)S_{QD}(\omega),
\end{equation}
\begin{equation}
 \Gamma_\nu = \int\!\dd\omega\, g_\nu^2(\omega)S_\nu(\omega)S_{QD}(\omega),
\end{equation}

\noindent where we have defined the QD spectrum (in absence of any cavity) as

\begin{equation}
S_\mathrm{QD}(\omega) = \int\!\dd\tau \,e^{-i\omega\tau}\braket{\sigma_+(\tau)\sigma_-(0)},
\end{equation}

The expression of the correlator $\braket{\sigma_+(\tau)\sigma_-(0)}$ is known and derive from an exact resolution of the independent boson model (see \cite{mahan}):

\begin{equation}
 \braket{\sigma_+(\tau)\sigma_-(0)} = e^{i\omega_{QD}\tau}e^{-\Phi(\tau)},
\end{equation}
\begin{multline}
 \Phi(\tau) = \int_0^{+\infty} \!\dd\omega_\mathbf{q} \rho_{ph}(\omega_\mathbf q) \left(\frac{M(\omega_\mathbf{q})}{\omega_\mathbf{q}}\right)^2\\
 \left[ N(\omega_\mathbf{q})(1-e^{i\omega_\mathbf{q}\tau}) + (N(\omega_\mathbf{q})+1)(1-e^{-i\omega_\mathbf{q}\tau}) \right],
\end{multline}

\noindent where $N(\omega_\mathbf{q}) = 1/(e^{\beta\hbar\omega_\mathbf{q}}-1)$ and $\omega^X_{QD} = 2\pi c/\lambda_X =  \omega_0 - \int\dd\omega_\mathbf q \rho_{ph}(\omega_\mathbf q) M(\omega_\mathbf q) /\omega_\mathbf q$ is the frequency of the exciton shifted by the so-called polaron shift. This expression can be rewritten

\begin{equation}
 \Phi(\tau) = \int_{-\infty}^{+\infty} \!\dd\omega\, f(\omega)(1-e^{-i\omega\tau}),
\end{equation}
\begin{equation}
 f(\omega) = \rho_{ph}(\omega)\left(\frac{M(\omega)}{\omega}\right)^2(N(|\omega|)+\theta(\omega)).
\end{equation}

\noindent The spectrum of the quantum dot is then given by

\begin{equation}
 \begin{split}
  S_{QD}(\omega) &= \int\!\dd\tau \exp\left[ \int\!\dd\omega\, f(\omega)\left( e^{-i\omega\tau} -1 \right) \right] e^{-i(\omega-\omega_{QD})\tau}\\
	&= 2\pi Z\delta(\omega_{QD}-\omega) \\ &+ Z\int\!\dd\tau\, \left[ \exp\left(\int\!\dd\omega\, f(\omega)e^{-i\omega\tau}\right) -1 \right]e^{-i(\omega-\omega_{QD})\tau}\\
	&= S_{QD}^{ZPL}(\omega) + S_{QD}^{PSB}(\omega),
 \end{split}
 \label{SQD}
\end{equation}

\noindent with $Z = e^{-\int\!\dd\omega\, f(\omega)}$. The first term in \eqref{SQD} corresponds to the zero phonon line, the second takes into account all orders of phonon assisted mechanisms. To acount for the finite lifetime of the exciton, the delta function in the ZPL spectrum can be replaced by a lorentzian. This artificial broadening has been used in figures 3., using the experimental width of the ZPL.\\
\indent Four relaxation channels for the QD can now be defined, the rate of each being 

\begin{equation}
\Gamma_{\nu}^{ZPL/PSB} = \int d\omega \rho_\nu(\omega) g^2_\nu(\omega) S_{QD}^{ZPL/PSB}(\omega).
\end{equation}

The effective fraction of emission coupled to the mode is

\begin{equation}
 \beta^{eff} = \frac{\Gamma^{ZPL}_\parallel + \Gamma^{PSB}_\parallel}{\Gamma_{tot}},
\end{equation}

\noindent where $\Gamma_{tot}$ is the total relaxation rate, \emph{i.e.} it involves the four relaxation processes mentioned above. We define $\beta^{ZPL}$ and $\beta^{PSB}$ the same way:

\begin{equation}
 \beta^{ZPL} = \frac{\Gamma^{ZPL}_\parallel}{\Gamma_{tot}}, \quad
 \beta^{PSB} = \frac{\Gamma^{PSB}_\parallel}{\Gamma_{tot}}.
\end{equation}

\noindent The effective Purcell factor is then given by

\begin{equation}
 \begin{split}
 F_P^{eff} &= \frac{\beta^{eff}}{1-\beta^{eff}} = \frac{\Gamma_\parallel^{ZPL}+\Gamma_\parallel^{PSB}}{\Gamma_\perp}\\
	&= \frac{\tau_X^0}{\tau_X} - 1,
 \end{split}
\end{equation}

\noindent where $\tau_X$ ($\tau_X^0$) is the decay time of the quantum dot in the pillar cavity (planar cavity). The usual nominal Purcell factor is

\begin{equation}
 F_P = \frac{\Gamma_\parallel^{ZPL}}{\Gamma_\perp^{ZPL}}.
\end{equation}

\section{Photon correlations in the cavity mode at high power}

Here we estimate $g_{C,C}^{(2)}(0)$ when the pump power is high and multiple excitons can be excited. In this regime, the cavity mode peak is fed by the exciton as well as the biexciton. We make approximation that the total spectrum of the radiative cascade is the sum of the spectrum of the biexciton--exciton transition and exciton recombination. Each line is weighted by the probability of each event. The spectrum of the biexciton--exciton transition is equal to the spectrum of an exciton recombination with a shifted ZPL frequency $\omega_{QD}^{XX} = 2\pi c /\lambda_{XX}$. Both the X and XX spectra are then computed using the model presented in last section. The relative weight of these two events is obtained using :

\begin{equation}
 \frac{I_{XX}}{I_X} = \frac{P_{XX}\cdot \beta_{XX}^{ZPL}}{P_X\cdot \beta_X^{ZPL}},
\end{equation} 

\noindent where $I_{X}$ ($I_{XX}$) is the measured intensity of the X (XX) peak, which is identified with the intensity of the zero phonon lines, $P_{X}$ and $P_{XX}$ are the probabilities of the two events and $\beta_{X}^{ZPL}$ and $\beta_{XX}^{ZPL}$ are the fractions of emission coupled to the mode computed with our model. The number of excitons excited inside the dot is assumed to follow a poissonian distribution. $P_X$ ($P_{XX}$) correspond to the probability that more than one (two) exciton is excited. It is then given by

\begin{equation}
 P_X = 1-e^{-\bar N},
\end{equation}
\begin{equation}
 P_{XX} = 1 - (1+\bar N)e^{-\bar N},
\end{equation}

\noindent where $\bar N$ is the mean number of excitons in the dot. The ratio of the last two equations gives

\begin{equation}
 \frac{\bar Ne^{-\bar N}}{1-e^{-\bar N}} + \frac{P_{XX}}{P_X} - 1 = 0,
\end{equation}

\noindent which can be numerically solved to determine $\bar N$, and then $P_X$ and $P_{XX}$.\\
\indent The coincidence probability is defined as

\begin{equation}
 g_{C,C}^{(2)}(0) = \frac{\displaystyle \int_{-T/2}^{+T/2}\!\dd t \int_{-T/2}^{T/2}\!\dd t\, \braket{a_1^\dag(t)a_2^\dag(t+\tau))a_2(t+\tau)a_1(t)}}
		{\displaystyle\left|\int_{-T/2}^{T/2}\!\dd t\, \braket{a_1^\dag(t)a_1(t)}\right|^2},
 \label{g2}
\end{equation}

\noindent where the ladder operator $a_i$ corresponds to the field in the two branches of the HBT interferometer filtered at the cavity frequency and $T$ is the period between two pulses. The numerator of \eqref{g2} corresponds to the probability that both the biexciton and exciton emit a photon in the cavity peak in a row, which we identify to $P_{XX}\cdot \beta_{XX}^{PSB}\cdot \beta_X^{PSB}$. The integral in the denominator is the total number of photons sent to the interferometer at the frequency of the cavity peak, given by $P_X\cdot \beta_X^{PSB} + P_{XX}\cdot \beta_{XX}^{PSB}$. Thus, the coincidence probability is

\begin{equation}
 g_{C,C}^{(2)}(0) = \frac{P_{XX}\cdot \beta_{XX}^{PSB}\cdot \beta_X^{PSB}}{(P_X\cdot \beta_X^{PSB} + P_{XX}\cdot \beta_{XX}^{PSB})^2}.
\end{equation}

\section{Case of a NV center in a cavity}

The environment tuning of the source can apply to other solid-state emitters. As an illustration, we study the case of a NV center in diamond, for which the PSB spreads over $100 \ nm$ and can  represent more than $90$ \% of the emission \cite{phononNV}. For extracting NV center single photon emission, various cavities are investigated like low volume photonic crystal cavities \cite{H1,englund} or ultra-small open-access microcavities \cite{NVT}. Our theoretical approach can be directly adapted to other emitters with a known spectrum. To model the single photon emission of a single NV into the optical mode of such a cavity, we compute a room temperature NV center emission spectrum using  the fit parameters given in \cite{albrecht}  (figure \label{nv}a). 

Figure \label{nv}b presents the fraction of emission into the mode $\beta^{eff}$  as a function of the detuning between the ZPL and the cavity mode, normalized to the cavity linewidth $\gamma_c$. The considered  cavity presents a Q factor of $250$ and various volumes, corresponding to various nominal Purcell factors ranging from  $F_P=10$ to $F_P=152$. The effective Purcell factor calculated when the cavity is resonant to the ZPL which ranges from $0.21$ to $3.32$ is also indicated. As shown here, to obtain a bright single photon source using a NV center, there is no need to tune the cavity resonance to the NV center ZPL. The fraction of emission into the mode can reach very high values (typically above $40$ \%) over a very large spectral range (around $50$ cavity linewidths) as long as the nominal Purcell factor $F_P$ is above $20$, a reasonable value considering the current technology state of the art \cite{H1,englund,NVT}.

\begin{figure}
 \centering
 \includegraphics[width=1.00\linewidth]{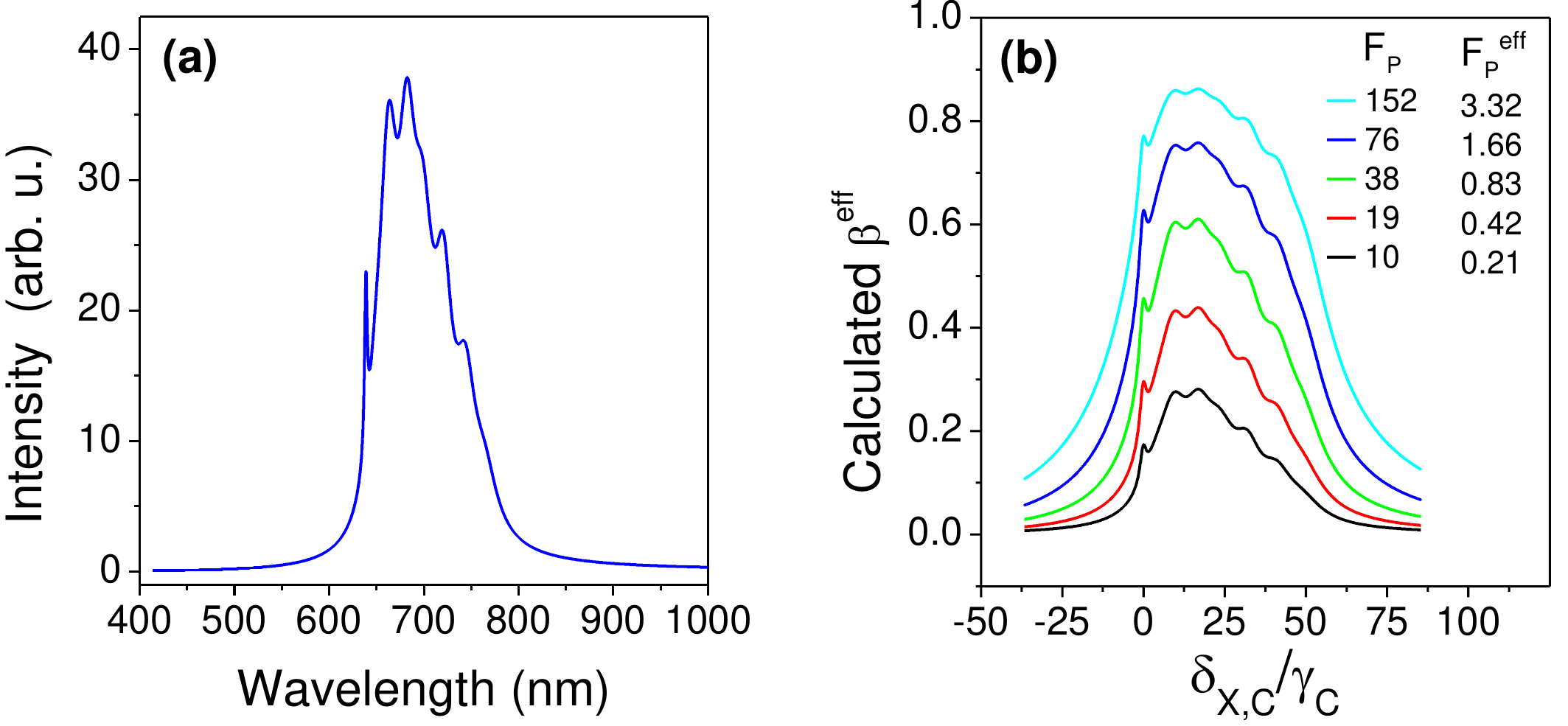}
 \caption{ (a) Computed spectrum of a NV center at room temperature plotted using the fit parameters given in \cite{albrecht}. (b)  NV center coupled to a cavity with Q=250. Calculated fraction of emission into the mode $\beta^{eff}$  as a function of the detuning between the cavity mode  and the ZPL normalized to the cavity linewidth.   $\beta^{eff}$ is calculated for different values of nominal Purcell factor $F_P$; the corresponding  effective $F_{P}^{eff}$ for the cavity resonant to the ZPL are also indicated.}
 \label{nv}
\end{figure}

\bibliographystyle{unsrt}

\end{document}